**Serkan Kucuksenel[1], Osman Gulseven[2]**
# ELECTORAL SYSTEMS AND INTERNATIONAL TRADE POLICY

*We develop a simple theoretic game a model to analyze the relationship between electoral systems and governments' choice in trade policies. We show that existence of international pressure or foreign lobby changes a government's final decision on trade policy, and trade policy in countries with proportional electoral system is more protectionist than in countries with majoritarian electoral system. Moreover, lobbies pay more to affect the trade policy outcomes in countries with proportional representation systems.*

*Keywords: electoral system; trade policy; lobby; design; optimum of Poreto.*

**1. Introduction.** In democracies, governments shape trade policy in response to not only the concerns of voters but also to lobbies pressure. Lobbies participate in political process to influence the outcomes. Hence, governments' choice of the implemented trade policy is affected by internal and external pressures. All parties in government should balance these 2 affects because they need the public support, necessary for reelection, but they also need the financial contributions that come from lobbies to finance everyday operations and campaign expenses.

---

[1] Corresponding author: Asst. Prof. of Economics, Middle East Technical University, Department of Economics, Ankara, Turkey.
[2] Asst. Prof. of Economics, Middle East Technical University, Department of Economics, Ankara, Turkey.

As Grossman and Helpman (1994) stated, 2 different approaches are prominent in the literature to explain the equilibrium in trade policy. The first one explicitly models the political competition between candidates and lobbies. Lobbies try to affect the outcome of elections by contributing to a party that promises the lobby more favorable trade policy. In the second approach, elections are not considered and the incumbent government maximizes the public support. For more on these different approaches see Hillman (1989). Our approach is in spirit with the second approach. We assume that there is an incumbent government in two different electoral systems and governments care about both public support and foreign lobby contributions.

Mansfield et al. (2000) analyze the relationship between regime type and trade policy. They find that aggregate trade barriers are lower within democratic pairs than within pairs composed of an autocracy and a democracy. Their argument depend on a legislature with the power to ratify executive proposals which is assumed to be common in democracies. However, they do not explain the reasons for different trade barriers within democracies.

In this paper, we present a simple theoretic game model that tries to capture the reasons for different trade policies between democracies. We believe that electoral systems are very important institutions in explaining this phenomenon. We label democracies with 2 electoral systems: proportional representation systems and majoritarian electoral systems. In the model, voters have single-peaked preferences on the tariff rate for an import good. We do not explicitly model the election process but we assume that there is an incumbent government which is formed by one party in the majoritarian electoral system and coalition of parties in the proportional representation system. In both economies, government care about both the citizens support and the monetary transfers from foreign lobbies. We assume that government needs monetary transfers for well functioning. Moreover, we assume that all agents are sincere. In both cases government has perfect information about the distribution of voters' preferences. The government bargains with a foreign lobby on tariff rate. We assume that the resulting outcome of the bargaining process is Pareto optimal, individually rational and symmetric. Then, we compare the equilibrium tariff rates and amount of monetary transfers between the 2 electoral systems.

The rest of the paper is organized as follows. In Section 2, we introduce the model for 2 different economies where economies are specified by their electoral systems. Section 3 contains our main results which basically state that tariff rates will be higher in proportional system. We conclude in Section 4 and discuss the implications of the results, sensitivity of the model to changes in assumptions and possible extensions of the model.

**2. The Model.** The model presented here has much in common with the Mayer (1984) and Kibris et al. (2003) framework. There is a finite set $N$ of voters. Let $X \subset \Re^+$ be a one dimensional trade policy (tariff rate) space. Each voter $i \in N$ is assumed to have single-peaked preferences on the tariff rate for an import good. The preference relation, $R_i$, is single-peaked if there is $p(R_i) \in \Re^+$, called the peak of $R_i$, such that for all $x_i$, $y_i$ in, $\Re^+$ $x_i < y_i \leq p(R_i)$ or $x_i > y_i \geq p(R_i)$ implies $y_i P_i x_i$. We assume that voters are not strategic so they are not aware of the bargaining process between lobby and government. Moreover, voters vote on tariff rate to declare to the government. We do not specify the election process but we assume that there is an incumbent government. Since all voters' preferences are single peaked on a single dimension, the median ideal tariff rate is a Condorcet winner and the social preference order under simple majority rule is transitive, with median standing in the highest order [Black (1948)]. Let $t_m > 0$ be median voter's ideal tariff rate.

In 2 party plurality systems, government is formed by one party. We are implicitly using Duverger's law which states that the plurality rule for selecting the winner of elections favors 2-party system. The government's decision is affected by public declaration which is median ideal tariff rate and a foreign lobby's most preferred tariff rate. Let $t_f$ be the lobby's most preferred tariff rate which is lower than that of median voter: $0 < t_f < t_m$.

The lobby grants favors in monetary terms to the government in return of a lower tariff rate. Let $M$ be the amount of monetary transfers. The government bargains with the lobby on the tariff rate and amount of transfers. The government's utility function may be written as:

$$G(t, M; t_m) = -(t_m - t)^2 + M. \qquad (1)$$

Hence, the government does not want to set tariff rate further from median ideal point unless it is compensated by the monetary transfers. The monetary transfers are not allocated to the voters. We assume that the transfers are necessary for well-functioning of the government or these are side transfers to the government. If we incorporate the allocation part to the model, we should assume that voters are sophisticated and as a result this process will increase the median voter's ideal tariff rate.

In proportional representation system, we know that party system is fragmented and there are usually more parties than those that elect by plurality or majority. As Bawn and Rosenbluth (2006) stated, coalition of many parties make systematically different policy choices than single party governments. We do not model the coalition formation process but we assume that government consists of a large party and a small party. For more on coalition formation in proportional representation systems see Austen-Smith and Banks (1988). Let $t_L^{PR}$ be the median ideal point of voters who support the large party and $t_L^{PR}$ be the median ideal point of voters who support the small party. The members' utility function of coalitional government may be written as follows:

$$G_L(t, M; t_L^{PR}) = -(t_L^{PR} - t)^2 + \alpha M, \qquad (2)$$

$$G_S(t, M; t_S^{PR}) = -(t_S^{PR} - t)^2 + (1 - \alpha)M. \qquad (3)$$

We assume that $t_L^{PR} > t_m > t_S^{PR} > 0$ and weight of the large party in the coalition is $\alpha$ such that $\alpha t_L^{PR} + (1 - \alpha) t_S^{PR} > t_m$. The monetary transfer the large party receives is proportional to its weight (or percentage of cabinet posts hold by the large party). Schofield and Laver (1985) show that approximately half of European coalition governments between 1945 and 1983 contain 2 parties and in most of the coalitions median party is not in the coalition. Hence, we believe that these assumptions are intuitively desirable.

The foreign lobby's preferences may be written as follows:

$$F(t,M) = \begin{cases} -(t-t_f)^2 - M, & \text{if } t > t_f \\ -M & \text{otherwise.} \end{cases} \quad (4)$$

Basically we assume that the tariff rates lower than $t_f$ do not make the lobby worse off but lower rates do not make the lobby better off. Hence, $t_f$ is the most preferred tariff rate of the foreign lobby. The second term (-M) summarizes the monetary benefits the lobby grants to the government.

In both electoral systems, the internal structure of the bargaining process between government and the foreign lobby is not specified. However, we assume that the outcome of bargaining between government and the lobby is Pareto optimal, individually rational and symmetric. Hence, the lobby and government have equal bargaining power and we use Nash bargaining solution concept. For more on this concept see Osborne and Rubinstein (1994). In proportional representation system, we use asymmetric Nash bargaining solution in which the ideal tariff rate of government is determined proportional to the weights of 2 parties in the government. Hence, the ideal tariff rate for coalitional government is $\alpha t_L^{PR} + (1-\alpha)t_S^{PR}$.

**3. The Main Results.** Our main result provides a comparison of implemented tariff rates and monetary transfers in different electoral systems.

*Theorem:* In the systems of proportional representation (PR), the implemented tariff rate $[(t)]_{PR}$ at equilibrium will be larger than the tariff rate ($t^*$) in those that elect by majority rule (SMD). Moreover, foreign lobbies transfer relatively more money to coalition governments.

*Proof:* From Pareto optimality, $t^*$ will satisfy the tangency condition,

$$\frac{\frac{\partial G(t,M;t_m)}{\partial t}}{\frac{\partial G(t,M;t_m)}{\partial M}} = \frac{\frac{\partial F(t,M)}{\partial t}}{\frac{\partial F(t,M)}{\partial M}} \Rightarrow 2(t^*-t_m) = 2(t^*-t_f) \quad (5)$$

Therefore, $t^* = \frac{t_m + t_f}{2}$ In PR, government's ideal point is $\alpha t_L^{PR} + (1-\alpha)t_S^{PR} > t_m$. By Pareto optimality,

$$t_{PR}^* = \frac{\alpha t_L^{PR} + (1-\alpha)t_S^{PR} + t_f}{2} \quad (6)$$

which implies $t^* < t^*_{PR}$. Given $t^*$, by individual rationality the lobby receives its disagreement payoff at $M^{max}$ and the government receives its disagreement payoff at $M^{min}$, where $M^{max} = (t_m - t_f)^2 - (t^* - t_f)^2 > 0$, and $M^{min} = (t_m - t^*)^2 > 0$. Since we assume that the government and the lobby have equal bargaining power, the bargaining process leads to the transfer $M^*$ where:

$$M^* = \frac{M^{max} + M^{min}}{2} = \frac{(t_m - t_f)^2 - (t^* - t_f)^2 + (t_m - t^*)^2}{2} \quad (7)$$

In PR, by individual rationality:

$$M_{PR}^* = \frac{M_{PR}^{max} + M_{PR}^{min}}{2} = \frac{\alpha t_L^{PR} + (1-\alpha)t_S^{PR} - t_f)^2 - (t_{PR}^* - t_f)^2 + (\alpha t_L^{PR} + (1-\alpha)t_S^{PR} - t_{PR}^*)^2}{2} \quad (8)$$

Therefore, $M^*_{PR} > M^*$.

**4. Conclusion.** This model tries to capture the reasons for different macroeconomic policies around the world. Our results suggest institutions might be an important factor in explaining these differences. We show that tariff rates will be higher in countries with proportional representation systems than those with majoritarian ones. The predictions of our model are also consistent with the empirical results. An empirical study by Rogowski and Kayser (2002) shows that prices of goods and services are higher in PR countries after controlling for other variables. Higher tariff rates imply goods at home country will be expensive since producers are protected from foreign competition and they can charge close to monopoly prices.

**References:**

*Austen-Smith, D., J. Banks* (1988). Elections, Coalitions, and Legislative Outcomes, American Political Science Review, 82, 405-422.

*Bawn, K., F. Rosenbluth* (2006). Coalition Parties versus Coalitions of Parties: How Electoral Agency Shapes the Political Logic of Costs and Benefits, American Journal of Political Science, 50, 251-265.

*Black, D.* (1948). On the Rationale of Group Decision Making, Journal of Political Economy, 56, 23-24.

*Grossman, G.M., Helpman E.* (1994). Protection for Sale, American Economic Review, 84: 4, 833-850.

*Hilmann, A.L.* (1989). The Political Economy of Protection, Chur: Harwood Academic Publishers.

*Kibris, A., O. Kibris, M.Y. Gurdal* (2003). Protectionist Demand in Globalization, Sabanci University, Turkey.

*Mansfield, E.D, H.V. Milner, B.P. Rosendorff* (2000). Free to Trade: Democracies, Autocracies, and International Trade, American Political Science Review, 94, 305-321.

*Mayer, W.* (1984). Endogenous Tariff Formation, American Economic Review, 74:5, 970-985.

*Osborne, M.J., A. Rubistein* (1994). A Course in Game Theory, New York, U.S.A.: The MIT Press.

*Rogowski, R., M.A. Kayser* (2002). Majoritarian Electoral Systems and Consumer Power: Price-Level Evidence from the OECD Countries, American Journal of Political Science, 46, 526-539.

*Schofield, N., M. Laver* (1985). Bargaining Theory and Portfolio Payoffs in European Coalition Governments 1945-83, British Journal of Political Science, 15:2, 143-164.